\documentstyle[twocolumn,prl,aps]{revtex}
\begin{document}
\draft 
\title{Quantum gambling using three nonorthogonal states 
%with reduced bias
 }
\author{Won-Young Hwang \cite{email}, and
Keiji Matsumoto \cite{byline}}
\address{IMAI Quantum Computation and Information Project,
ERATO, Japan Science Technology Corporation,
Daini Hongo White Bldg. 201, 5-28-3, Hongo, Bunkyo,
Tokyo 133-0033, Japan}
\maketitle

\begin{abstract}
We provide a quantum gambling protocol using three (symmetric)
 nonorthogonal states. The bias of the proposed protocol is less 
than that of previous ones, making it more practical. 
We show that the proposed  scheme is secure against
nonentanglement attacks. The security of the proposed scheme
against entanglement attacks is shown heuristically.
\end{abstract}


\narrowtext

\section{introduction}
%%%%%%%%%%%%%%%%%%%%%%%%%%%%%%%%%
Unforgeable quantum money proposed by Wiesner 
\cite{wies} opened the field of `quantum cryptography'. 
The most successful of the quantum cryptographic 
protocols is Bennett and Brassard (BB84) quantum key
 distribution (QKD) protocol \cite{bene}, whose unconditional
 security was proved more than a decade later \cite{maye}. 
Since another very useful ingredient in cryptographic tasks is the
 bit commitment, there has been much effort to find 
an unconditionally
 secure quantum bit commitment protocol. However, it turns out
 that no such thing exists \cite{lo,may2}. This fact 
motivated the search for a slightly weaker protocol,
 quantum coin tossing. However, it turns out that the ideal
 quantum coin tossing protocol also does not exist \cite{lo2}.
 It is still
 an open question whether almost ideal quantum coin tossing exists
 or not \cite{amba}. However, it was found that there exists
 a quantum gambling protocol that is weaker than quantum
 coin tossing \cite{gold}. 

We can say that the quantum money and the BB84 protocol are
 based on a basic property of quantum mechanics, the
 no-cloning theorem \cite{woot,diek}. Another closely related but
 different property in quantum mechanics is that nonorthogonal 
quantum states cannot be distinguished with certainty \cite{yuen}. 
It is
 interesting to search for quantum protocols utilizing this property.
 Bennett's later QKD scheme indeed utilizes this property
 \cite{ben2}. Recently, Hwang {\it et al.} gave a quantum gambling
 scheme that utilizes this basic property \cite{hwan}. 

The two
 quantum gambling protocols \cite{gold,hwan} are not ideal in
 the sense that there is a bias $\delta >0$: It is an unfair game by
 the amount of the bias $\delta$. That is, for each round of the
 game the expectation value of one party's gain is given by the
 bias $\delta$.  However, since the bias $\delta$ is proportional
 to $1/\sqrt{R}$ where $R$ is the money penalty, the bias $\delta$
 can be made negligible by making $R$ very large in both 
schemes \cite{gold,hwan}. 

In this paper, we provide a quantum gambling protocol using three
 nonorthogonal states. In the proposed scheme, two participants
 Alice and Bob can be regarded as playing a game of making
 guesses at the
 identities of quantum states that are in one of three
given nonorthogonal states: If Bob makes a correct (incorrect)
guess at the identity
of a quantum state that Alice has sent, he wins (loses).
We show that the proposed  scheme is secure against
 non-entanglement attacks. The security of the proposed scheme
 against entanglement attacks is shown heuristically. However,
 since the idea behind the proof is simple, we believe that 
a rigorous one will be found as in the case of the QKD 
\cite{maye,shor,lo3}. The advantage of the proposed scheme over
 previous ones is that the bias $\delta$ is proportional to $1/R$. 
We discuss this advantage.
%%%%%%%%%%%%%%%%%%%%%%%%%%%%%%%%%%%
\section{quantum gambling using three nonorthogonal states}
%%%%%%%%%%%%%%%%%%%%%%%%%%%%%%%%%%%
Let us now describe the three symmetric nonorthogonal states to
 be used in the protocol. 
Let $\{p_i, |i\rangle \langle i| \}$
denote a mixture of pure states $|i\rangle \langle i|$
with relative frequency $p_i$ with $\sum_{i} p_i = 1$. 
$\rho= \sum_i p_i |i\rangle \langle i|$ is a density operator
  that corresponds to the mixture $\{p_i, |i\rangle \langle i| \}$.  
Any pure quantum bits (qubits) $|i\rangle \langle i|$ can be
 represented by a (three-dimensional Euclidean) Bloch vecter
 $\hat{r}_i$ as $|i\rangle \langle i|=(1/2)({\bf 1}+ \hat{r}_i
 \cdot {\bf \vec{\sigma}})$ \cite{niel}. Here ${\bf 1}$  is the
 identity operator, 
${\bf \vec{\sigma}}= (\sigma_x, \sigma_y, \sigma_z)$,
and $\sigma_x, \sigma_y, \sigma_z$ are the Pauli operators. 
The Bloch vectors of the three nonorthogonal states $|a\rangle$,
 $|b\rangle$, and $|c\rangle$ are in the same plane and make 
an angle $2\pi/3$ with one another to be symmetric. 
Here we adopt $|a\rangle= |0\rangle$, 
$|b\rangle= 1/2 |0\rangle + \sqrt{3}/2 |1\rangle$,
 and $|c\rangle= 1/2 |0\rangle - \sqrt{3}/2 |1\rangle$, where 
$|0\rangle$ and $|1\rangle$ denote two mutually
 orthogonal states of a qubit as usual.  
 
Let us now give the protocol.\\
(1) Alice randomly chooses one among the three nonorthogonal
 states
$|a\rangle$, $|b\rangle$, and $|c\rangle$, and sends it to Bob.\\
(2) On the qubit he receives, Bob performs an optimal
 measurement, that is, a measurement by which he can obtain the
 maximal probability $p$ of correctly guessing the identity of the
 qubit. \\  
(3) On the basis of the measurement's results, 
he makes a guess at
 which one the qubit is and annouces it to Alice. \\
(4) If he made a correct (incorrect) guess, Alice announces he
 has won (lost).\\
(5) When Bob has won, Alice gives
him one coin.  When he has lost,
Bob gives her $p/(1-p)$ coins. 

However, after the first step,
Bob follows the following steps $6-9$ instead of 
steps $2-5$,
at randomly chosen instances with a rate $r$
($0<r \ll 1$).\\
(6) Bob performs no measurement on the qubit and
 stores it.\\
(7) He announces his randomly chosen guess at the
 identity of the qubit.\\
(8) Step 4 is repeated.\\
(9) In the previous step,
Alice has actually revealed which one
she chose to tell him the qubit is
(regardless of her honesty).
When it is $|\alpha \rangle$ ($\alpha= a,b,c$),
Bob performs $\hat{S}_\alpha$.
($\hat{S}_\alpha$ is an orthogonal measurement
that is composed of two projection operators
$|\alpha \rangle \langle \alpha|$ and 
$|\alpha^{\prime}\rangle \langle \alpha^{\prime}|$. 
Here $|\alpha^{\prime}\rangle$ is a normalized state that is
 orthogonal to $|\alpha \rangle$.)
If the outcome is $|\alpha^{\prime} \rangle$, 
Bob announces that he
performed $\hat{S}_\alpha$ and got $|\alpha^{\prime}\rangle$
as an outcome. Then Alice must give him $R$
($\gg 1 $) coins.
If the outcome is $|\alpha\rangle$, Bob says nothing about
which measurement he performed and follows step 5. $\Box$ 

As in the two-state scheme \cite{hwan}, it is important in step 2
 for Bob to perform the optimal measurement that assures
 maximal probability $p$ of correctly guessing the identity of the
 qubit, in order to assure his maximal gain. The optimal
 measurment for the three nonorthogonal states
 $|\alpha\rangle$ was recently given \cite{ande}. It is 
a positive operator valued measurement (POVM) \cite{pere}
 whose component operators are, interestingly, proportional to
 the three operaters $|\alpha\rangle \langle\alpha|$
 \cite{ande}.
That is, they are $ (2/3) |a \rangle \langle a|$,
$ (2/3) |b \rangle \langle b|$, and $ (2/3) |c \rangle \langle c|$. 
Now it is easy to see that the maximal probability 
$p$ is $2/3$.
%%%%%%%%%%%%%%%%%%%%%%%%%%%%%%%%%%
\section{security of the protocol}
%%%%%%%%%%%%%%%%%%%%%%%%%%%%%%%%%%
Now let us show how each player's average gain is assured. 
(Here we repeat the corresponding part of Ref. \cite{hwan} in
 a slightly varied form.) 
 
First it is clear by definition that Bob can do nothing better than
 performing the optimal measurement,
as long as Alice prepares the specified qubits.
In the protocol, the numbers of coins that Alice and Bob pay
are adjusted so that no one gains when Bob's
win probability is $p$. Thus Bob's gain $G_B$ cannot be greater
than zero, that is, $G_B \leq 0$.

Next let us consider Alice's strategy.
As noted above, we first show the security against
Alice's nonentanglement attacks.
Roughly speaking, Alice can do nothing but prepare
the given states $|\alpha\rangle$ and honestly
tell Bob the identity of the state later.
Otherwise she must pay $R$ ($\gg 1 $) coins to him sometimes,
making her gain negative. Let us consider this more precisely.
In the most general nonentanglement attacks,
Alice randomly generates each qubit in a state $|i\rangle$
with a probability $p_i$.
Here the $|i\rangle$'s are arbitrarily specified states
of qubits,
$i=1,2,. . .,N$ and $\sum_i^N p_i=1$. However, since Bob has
no information about which $|i\rangle$ Alice has selected at each
instance, his treatments of the qubits actually become equal for all
qubits. Thus it is sufficient to show the security for a qubit in 
an arbitrary state. Let us denote the angles that the Bloch vector of
 a state $|i\rangle$ makes with those of $|\alpha\rangle$ as
 $\theta_\alpha$. 
At ramdomly chosen instances with a rate $r$, Bob checks Alice's
 claim by measuring 
$\hat{S}_\alpha$ when the claim is that the state is 
$|\alpha\rangle$ (the steps $6-9$).
If the measurement's outcomes are $|\alpha^{\prime}\rangle$,
the claim is proved wrong. Then Alice must give
Bob $R$ coins. The probability that a state $|i\rangle$ is checked is
 $|\langle \alpha^{\prime}| i \rangle|^2=
 1- \cos^2(\theta_\alpha/2)$ in the case when the checking
 measurement
$\hat{S}_\alpha$ is performed. Thus one term in Alice's gain
 $G_A$ is $-rR[1- \cos^2(\theta_\alpha/2)]$ where $rR$ is set to
 be much larger than $1$. 
Now it is simple to see that Alice should prepare  
only states that are highly nonorthogonal to one of the
 $|\alpha\rangle$'s. Thus one of the $\theta_\alpha$ is very
 small. Otherwise, Alice's gain $G_A$ will be dominated by highly 
negative term $-rR [1- \cos^2(\theta_\alpha/2)]$ in any case. 
Similarly we can see that  she should claim the prepared state to
 be the one that is nearest to it. 
Here it should be noted that 
we should take into account the fact that
Alice obtains partial information about whether Bob has
performed the measurement or not, due to Bayes' rule. 
However, Alice still cannot increase her gain as long as the $R$
is large enough, because she cannot
 be confident that Bob has already performed the measurement.
Let $f_u$ be Alice's estimation of the probability that
Bob did not perform the measurement.
With no information, $f_u $ is $r$.
However, Bob's announced guess gives her
partial information about his measurement's result
if he performed it. This information can be used to
make a better estimate of $f_u$. For example,
in the case where Alice sends $|\alpha \rangle$
and Bob performs the optimal measurement,
we obtain using Bayes' rule that 
$f_u= (r/3)/[(r/3)+(1-r)(2/3)]$
when his guess is $|\alpha\rangle$.
However, it is clear that $f_u \geq r/3$:
when Bob did not perform the measurement,
he simply guesses it with equal probabilities
regardless of what he received.
Thus, by Bayes' rule, Alice can see that there remains 
a probability greater than $r/3$ that Bob did not perform
the measurement. The relation $f_u \geq r/3$ also holds
for the entanglement attacks, since it is satisfied for any
$|\alpha \rangle$.

Now let us consider a state $|i\rangle$ that satisfies 
the requirements
 $\theta_a \sim 0$, $\theta_b \sim 2\pi/3$, 
and $\theta_c \sim 2\pi/3$, without loss of generality.
The probability $P_C$  that Bob makes a correct guess is given
 by  $P_C= (2/3) \cos^2(\theta_a/2)$.
 That for an incorrect one is given by 
$P_I= 1-P_C $. Alice's gain is $-1$ ($2$) when Bob makes
 a correct (incorrect) guess. Let us denote Alice's gain
 $G_A^n$ ($G_A^c$) in the case of the normal (checking) steps.
 Alice's total gain is given by $G_A= (1-r)G_A^n+ r G_A^c$. 
Alice's gain $G_A^n$ in the case of the normal steps can be
obtained as
\begin{eqnarray}
\label{a}
G_A^n 
&=& (-1)(2/3) \cos^2(\theta_a/2) + 2 \{1- (2/3)\cos^2(\theta_a/2)\}
\nonumber\\
&=& 2\{1- \cos^2(\theta_a/2)\}.
\end{eqnarray}
Alice's gain $G_A^c$ in the case of the checking steps 
(when Alice claims that the sent qubit is $|a\rangle$) is given by
\begin{eqnarray}
\label{b}
G_A^c 
&=&  -R [1- \cos^2(\theta_a/2)]
    + \cos^2(\theta_a/2).
\end{eqnarray}
The second term in the right-hand side of Eq. (\ref{b}) is
because Bob makes a random guess without performing the 
optimal measurement in the checking steps and thus it is
 disadvantageous for him.
Then we can obtain that
\begin{eqnarray}
\label{c}
G_A 
&=& (1-r) 2\{1- \cos^2(\theta_a/2)\}-
rR \{1-\cos^2(\theta_a/2)\} 
\nonumber\\
&& + r \cos^2(\theta_a/2)
\nonumber\\
&=& \{2-r(R+2)\} \{1-\cos^2(\theta_a/2)\} 
              +r \cos^2(\theta_a/2). 
\end{eqnarray}
Here it is easy to see that if $r(R+2)\gg 2$ the optimal choice for Alice is that $\theta_a= 0$. Then the maximal gain for Alice is given by $G_A^{max}= r$. If we determine the values of $r$ and $R$ such that they satisfy the relation $r(R+2)=k \gg 2$ ($k$ is a constant), Alice's maximal gain or the bias $\delta$ is $r$. Thus {\it the bias $\delta$ is proportional to 1/R}.
The basic reason for this advantage is that the measured 
states $|\alpha\rangle$ coincide with the elements of the
 optimal POVM in the proposed scheme.
Alice could increase her gain $G_A^n$ for the normal steps by
 increasing $\theta_a$ in both two- and three-state schemes but 
with the following difference.
In the three-state (two-state) scheme,
$G_A^n$ increases with the second (first) order of $\theta_a$ 
while the probability to be checked increases with the
second order of $\theta_a$. 

Let us heuristically show the security against Alice's entanglement
 attacks. In entanglement attacks, she does not send a separate
 state but sends qubits that are entangled with some other qubits
 she preserves. If she can change Bob's state $\rho_B$ as 
she likes, she can always win. 
The basic idea is that she cannot do so even in entanglement
attacks. 
Instead, by appropriately choosing her measurement, Alice
can generate at Bob's site any ensemble
$\{p_i, |i\rangle \langle i| \}$ satisfying
$\sum_i p_i |i\rangle \langle i|= \rho_B$ (the theorem of 
Hughston, Jozsa, and Wootters) \cite{hugh}.
Let
$\rho_B= (1/2)({\bf 1}+ \hat{r} \cdot {\bf \vec{\sigma}})$.
Since $\rho_B= \sum_i p_i |i\rangle \langle i|$ and
$|i\rangle \langle i|=
(1/2)({\bf 1}+ \hat{r}_i \cdot {\bf \vec{\sigma}})$ where
$\hat{r}_i$ is the corresponding Bloch vector, we have
$(1/2)({\bf 1}+ \hat{r} \cdot {\bf \vec{\sigma}})=
(1/2)({\bf 1}+ [\sum_i p_i\hat{r}_i]
\cdot {\bf \vec{\sigma}})$ and thus
\begin{equation}
\label{d}
\hat{r}= \sum_i p_i\hat{r}_i.
\end{equation}
Therefore, for a given $\rho_B$
whose Bloch vector is $\hat{r}$,
Alice can prepare at Bob's site any
mixture $\{p_i, |i\rangle \langle i| \}$
as long as its Bloch
vectors $\hat{r}_i$ satisfy Eq. (\ref{d}).
However, if Alice always performs a given measurement,
the entanglement attacks reduce to the
nonentanglement attacks: The outcomes of measurements on
entangled pairs do not depend on the temporal
order of the two participants' measurements. So
we can confine ourselves to the case where Alice measures
first. Then the attack reduces to a nonentanglement
attack where Alice generates
$|i\rangle$ with probability $p_i$.
The only thing that Alice can do to utilize the
entanglement is to choose her measurements according to
Bob's announced guesses. However,
the checking steps also prevent Alice from increasing her
gain: She must choose the measurement that gives some mixture
$\{p_i, |i\rangle \langle i| \}$ at Bob's site such that each
$\hat{r}_i$ is the same as one of the Bloch vectors of the three 
nonorthogonal states $|\alpha\rangle$. This is because any vector
 $\hat{r}_i$ that deviates from those of the $|\alpha\rangle$'s will
 decrease Alice's gain due to the checking steps, involving 
 a negative term containing $rR$.
Therefore, Alice has no freedom in the
choice of measurements but a given choice. Thus the attack
 reduces to
 nonentanglement attacks for the reasons noted above. 
 %%%%%%%%%%%%%%%%%%%%%%%%%%%%%%%%%
\section{discussion and conclusion}
%%%%%%%%%%%%%%%%%%%%%%%%%%%%%%%%%
Let us discuss the advantage of the proposed scheme.
The problem of quantum gambling schemes is that Alice can
 claim that the error is due to noise or decoherence in
 the quantum 
channel, whenever it is checked and thus she must pay $R$ to
 Bob. 
This problem cannot be clearly solved even if quantum error
 correcting codes \cite{niel} are successfully implemented
  because a small amount of error always remains.
The solution to this problem is that Bob aborts the whole protocol
 if the error rate claimed by Alice is greater than the expected
 residual error rate.
%For this purpose, the number of errors should be large enough for
%the estimation of the error rate. 
However, Bob should actually accept his loss which amounts to
 the product of the number of errors and $R$, until data for 
a sufficient number of errors accumulate. Thus it is hard for Bob 
to do so when $R$ is too large. However, in the previous 
schemes (proposed scheme), we have that $R\sim 1/\delta^2$
 ($R\sim 1/\delta$), namely, for a given bias the value of $R$ of
 the proposed scheme is less than that of the previous schemes
 by a factor of $1/\delta$. Therefore we can say that the 
proposed scheme is more practical than previous ones.  

In conclusion, we provided a quantum gambling protocol using three 
(symmetric) nonorthogonal states. We showed that the proposed 
 scheme is secure against nonentanglement attacks. The
 security of the proposed scheme against entanglement attacks
 was shown heuristically. The advantage of the proposed scheme
 over previous ones is that the bias $\delta$ is proportional to
 $1/R$. We discussed its practical advantage.

\acknowledgments
We are very grateful to Professor Hiroshi Imai and the
Japan Science Technology Corporation for financial support. 
We are also very grateful to Dr. Alberto Carlini for helpful 
discussions.

\end{document}